# NUCLEI WITH TETRAHEDRAL SYMMETRY

J. DUDEK[1)], J. DOBACZEWSKI[2)], N. DUBRAY[3)], A. GÓŹDŹ[4)], V. PANGON[1)]
and N. SCHUNCK[5)]

[1] *Institut Pluridisciplinaire Hubert Curien, Departement de Recherches Subatomiques and Université Louis Pasteur, Strasbourg I, 23 rue du Loess, F-67037 Strasbourg, France*
*Jerzy.Dudek@ires.in2p3.fr, Vincent.Pangon@ires.in2p3.fr*

[2] *Institute of Theoretical Physics, Warsaw University, Hoża 69, PL-00681 Warsaw, Poland*
*Jacek.Dobaczewski@fuw.edu.pl*

[3] *CEA/DAM Ile de France, DPTA/Service de Physique Nucléaire, BP 12, F-91680 Bruyères-le-Châtel, France*
*Noel.Dubray@cea.fr*

[4] *Zakład Fizyki Matematycznej, Uniwersytet Marii Curie-Skłodowskiej , pl. Marii Curie-Skłodowskiej 1, PL-20031 Lublin, Poland*
*Andrzej.Gozdz@umcs.lublin.pl*

[5] *Departamento de Fisica Teorica, Universidad Autonoma de Madrid SP-28049 Cantoblanco, Madrid, Spain*
*Nicolas.Schunck@uam.es*

We discuss a point-group-theory based method of searching for new regions of nuclear stability. We illustrate the related strategy with realistic calculations employing the tetrahedral and the octahedral point groups. In particular, several nuclei in the Rare Earth region appear as excellent candidates to study the new mechanism.

## 1. Introduction

The problem of nuclear stability and of examining its various forms has become one of the hot issues in nuclear structure physics in view of the progress in construction of the large scale experimental facilities such as, e.g., GSI-Darmstadt (Germany), GANIL (France), or RIKEN (Japan), focusing on the properties of exotic nuclei. Some new avenues in this domain of research correspond to the exploitation of point-group symmetries – so far almost completely neglected in nuclear structure calculations, although extensively employed in other domains, notably in molecular physics. The arguments that deformed potentials may give rise to the Fermion-level degeneracies higher than 2 in realistic $N$-body systems like metallic clusters were formulated in Ref.[1]. The first results of realistic calculations applied to heavy nuclei of $A \sim 230$ region, predicting the presence of the hypothetical tetrahedral minima together with exotic four-fold degeneracies, can be found in Ref.[2]. In the latter reference, the nuclear tetrahedral 'magic numbers' Z=56, 64, 70, and 90 and N=90, 112, and 136 were for the first time predicted. The tetrahedral symmetry instability has been studied in the light nuclei using self-consistent Hartree-Fock method.[3] A more complete survey of the tetrahedral magic numbers together with the group theoretical arguments about the research of the new forms of nuclear stability can be found in Ref.[4]. Some progress in relation to the description of the collective nuclear rotation using the methods adapted from those used in the molecular physics has been presented,[5] followed by a discussion of the problem of redundant variables in the nuclear quantum rotors.[6] A possibility of appearance of the pure *octahedral symmetry* nuclear configurations has also been discussed.[7] The possibility that some nuclei may be tetrahedral-symmetric in their ground-states[8] has been put forward in 2004. Very recently the arguments that the tetrahedral symmetry has already been seen in experiments were given,[9] while the first systematic Hartree-Fock-Bogolyubov calculations using different forces have been presented in Ref.[10]. Finally, the first investigation of the possible tetrahedral nuclear vibrations using the Generator Coordinate Method can be found in Ref.[11].

## 2. Symmetry Arguments - Qualitatively

The main arguments related to exploiting the point-group symmetries and the implied group structure can be summarized as follows. Firstly, let us remark that the stability of an $N$-body quantum system depends on the excitation energy required to transform its ground-state into the first excited state. Within the mean-field approach, one of the most successful realistic methods in nuclear structure physics, such an excitation energy will be the larger the bigger the single-particle gap at the Fermi energy. Thus the problem of searching for the strongest nuclear stability can be approached, within the mean-field techniques, by the search for the possibly largest gaps in the single-particle spectra. As a consequence, the next step in our considerations must address the mechanisms that can possibly help in looking for the maximum gaps as functionals of nuclear shapes and associated symmetry. There are two such mechanisms that influence the solution: one is related to the saturation of the nuclear forces while another one, combined with the former, has to do with the point-group symmetries, the implied number of irreducible representations, and the dimensions of the irreducible representations in question.

The saturation property of the nuclear forces implies, among others, that the depth of the nuclear mean-field potential well, say $V_0$, remains, to an approximation, independent of the particle number, neither does it seem to depend on the nuclear deformation. Its depth of the order of $V_0 \in [-60, -50]$ MeV can be considered as an average 'universal' estimate holding as the zero'th order approximation for all nuclei throughout the Periodic Table. Consequently, for a fixed particle number, say $N = N_p$ or $N_n$, the number of protons or neutrons, a rough estimate gives the density, $\rho_\lambda$, of single-particle proton and/or neutron levels in the vicinity of the Fermi energy as proportional to the ratio $N/V_0$, and the average level spacing $d \sim 1/\rho_\lambda \sim V_0/N$. This zero-order estimate does not take into account symmetries. To clarify what is meant here, first consider two parities of the discussed states. If the numbers of states in both parities are $N^+$ and $N^-$, the average level spacings in both parities separately will be $d^{(+)} \sim V_0/N^+$ and similarly $d^{(-)} \sim V_0/N^-$. With the increasing nuclear mass $|N^+ - N^-| \to 0$ and therefore $N^+ \sim N^- \sim N/2$: the average level spacings in the two parities will be twice as large as compared to the previous estimate without symmetries taken into account. Obviously what interests us in the present context is the average level spacing within *each symmetry separately.* This is because by combining relatively large gaps from both symmetries we may possibly arrive at a large common gap at the Fermi energy and therefore at the increased stability. Thus the necessary (although not sufficient) condition reads: *Maximize the gaps in each symmetry separately* – this may, although does not need to, lead to the common absolute large gaps in the single-particle spectra.

Suppose now that there exists a symmetry group $G$ of the mean field which has several, say $g$, irreducible representations $R_r$; $r = 1, 2, \ldots g$. An important property of the repartition of single-particle solutions over irreducible representations is that the energy levels belonging to *distinct* irreducible representations never repel each other (they cross freely in function of the symmetry-preserving deformations), while there are in general no crossings allowed within a given irreducible representation. Denote the numbers of levels corresponding to various irreducible representations by $n^{(r)}$; we find that there will be $r$ estimates of the average level spacings according to symmetry as given by $d^{(r)} \sim V_0/(N^{(r)})$. In general, the numbers $N^{(r)}$ corresponding to different irreducible representations will be different. Let us suppose, nevertheless, as a heuristic hypothesis, that the repartition of the original $N$ levels over the $g$ irreducible representations is approximately uniform in which case $N^{(r)} \approx N/g$. Let $g = 6$ as an example. In such a case there will be a factor of 6 gain in terms of the average level spacings as compared to the originally considered situation. We thus conclude that the mean-field Hamiltonian invariant with respect to a point group symmetry with large number of irreducible representations has, on average, increased chances to lead to big gaps in the single-particle spectra, thus possibly leading also to the new class of nuclear stability and/or isomerism.

It is clear that the average level spacing can only serve as the first guide-line. The fluctuations around these averages are expected (and shown by numerical calculations with the realistic potentials) to be significant. This is precisely the effect sought: the bigger the deviation the better the chance to arrive at large gaps in the spectra. At this point, the argument about the saturation of the nuclear forces and the nearly constant potential depth can be helpful. To illustrate this aspect consider a fluctuation consisting in an *increase of the local level density* at a certain energy range by a certain factor. Because of the constant depth of the potential this automatically implies that the local density in some other energy range of the spectrum must *decrease* and the corresponding gaps become larger, i.e., exactly what we are looking for.

Another mechanism of importance related to the group-theoretical considerations corresponds to the dimensionality of the irreducible representations. More precisely, if in addition to the large number of irreducible representations, some of them have large dimensions, say $d_r$, the implied single-particle level energies will appear as $d_r$-fold degenerate. The presence of high degeneracies increases, on average, the single-particle level spacings thus acting in the desired direction. At this point it will be appropriate to recall that there are two types of point group symmetries sometimes referred to as 'simple' and 'double', the former applying to the 3D geometry of, e.g., macroscopic bodies, while the latter applying to systems and Hamiltonians composed of fermions. It is well known that the dimensions of the irreducible representations of the double point groups can be *at most* equal to 4. Consequently, the term 'large dimension $d_r$' may imply in the best case $d_r = 4$. As it turns out, the octahedral double-group is characterized by $g = 6$ irreducible representations, two of them with the dimension equal to 4, thus representing a good candidate case. The mathematically related (see below) tetrahedral double point group has $g = 3$ and one four-dimensional irreducible representation.

On the basis of the above considerations, we would like to suggest that the point-group-symmetry guided research of nuclear stability appears as an interesting hypothesis, so far exploited to a very limited extent only. All the necessary details connected to the mathematical background of the point-group theory and related irreducible representations are well known, which facilitates the applications considerably. While a detailed study in this direction is in progress, in what follows we concentrate on two types of (related) symmetries, namely, on the tetrahedral and octahedral one. These two appear promising according to the calculations of the total potential energies, as illustrated further in this paper.

## 3. Realistic Mean-Field Hamiltonian and Method

An ideal tool to study the point-group symmetries and their impact on the implied properties of the nuclear single-particle spectra are *non-selfconsistent* realizations of the mean-field approach with parametrized potentials, as it will be illustrated in Sect. 5 below. One of the very successful approaches of this kind is provided by the Hamiltonians with the deformed Woods-Saxon type potentials with its universal[a] parametrization that has been in use over many years by now. The corresponding mean-field Hamiltonian, $\hat{H}_{\rm mf}$, has the form,

$$\hat{H}_{\rm mf} = \hat{t} + V_c(\vec{r}\,) + V_{so}(\vec{r}, \vec{p}, \vec{s}), \tag{1}$$

[a]The adjective 'universal' refers to the fact that the three parameters of the central potential, viz. depth $\bar{V}$, radius $r_0$, and diffusivity $a$, and the corresponding parameters of the spin-orbit potential can be chosen once for all nuclei of the Periodic Table providing a very good description, on average, of the single-particle spectra from light to heaviest nuclei. For more details see below.

with the central potential defined by

$$V_c(\vec{r}) \equiv \frac{\bar{V}}{1+\exp[\mathrm{dist}_\Sigma(\vec{r})/a]} + \frac{1}{2}(1+\tau_3)\, V_{Coulomb}(\vec{r}). \tag{2}$$

Above, $\bar{V} \equiv V_0\,[1 \pm \kappa(N-Z)/(N+Z)]$ (signs: "+" for protons, "-" for neutrons); $V_0$, $\kappa$, and $a$ are adjustable parameters and $\tau_3$ denotes the third component of the isospin. The symbol $\mathrm{dist}_\Sigma(\vec{r})$ denotes the distance between the current point position $\vec{r}$ and the nuclear surface $\Sigma$ defined in terms of the spherical harmonics $Y_{\lambda\mu}(\vartheta,\varphi)$ by

$$\Sigma: \quad R(\vartheta,\varphi) = R_0 c(\{\alpha\})\left[1 + \sum_{\lambda=2}^{\lambda_{max}} \sum_{\mu=-\lambda}^{\lambda} \alpha^\star_{\lambda\mu} Y_{\lambda\mu}(\vartheta,\varphi)\right]. \tag{3}$$

Above, $R_0 \equiv r_0\, A^{1/3}$ is the nuclear radius parameter and the function $c(\{\alpha\})$ takes care of the nuclear constant volume that is kept independent of the nuclear deformation. Note that there are effectively three parameters of the central potential for protons and three for neutrons; they are denoted as $\bar{V}, r_0$, and $a$ – the central potential depth, radius, and diffusivity parameters, respectively. It is sometimes convenient to introduce an alternative representation that replaces $\bar{V}$ for the protons and a similar parameter for the neutrons by a suitably chosen set of two parameters. In our case these are $V_0$ and $\kappa$ as introduced just below Eq. (2).

The spin-orbit potential has the usual form,

$$V_{so}(\vec{r},\vec{p},\vec{s}) \equiv \lambda_{so}[\nabla V(\vec{r}) \wedge \vec{p}] \cdot \vec{s}, \tag{4}$$

where $V(\vec{r})$ is another Woods-Saxon type deformed potential that differs from the analogous term in the central potential by the numerical values of the adjustable constants: here $\lambda_{so}$, $r_0^{(so)}$, and $a_{so}$.

Beginning with the single-particle spectra obtained by diagonalization of the above Hamiltonian within the Cartesian harmonic oscillator basis we apply the macroscopic-microscopic method of Strutinsky using the macroscopic Yukawa-plus-exponential approach. The formalism that we use here was presented in details in Refs.[12,13].

## 4. Nuclear Surfaces Invariant Under Symmetry Point-Groups

We are going to consider the nuclear surface equation written down in the form of the expansion in terms of the spherical harmonics, cf. Eq. (3). We wish to write down the nuclear mean-field Hamiltonian with the deformed Woods-Saxon and spin-orbit potentials, Eqs. (1) and (2), that is invariant under all the operations $\hat{g}$ of a given symmetry point-group $\mathbf{G}$. Here $\hat{g}$ denotes any point-group symmetry operation such as finite-angle rotations, proper or improper[b], plane reflections, and possibly inversion. The condition of invariance implies, by definition, that under the action of any group element $\Sigma \xrightarrow{\hat{g}} \Sigma' \equiv \Sigma$. The latter can be written down as

$$\sum_{\lambda=2}^{\lambda_{max}} \sum_{\mu=-\lambda}^{\lambda} \alpha^{\star}_{\lambda\mu} \, [\hat{g} \, Y_{\lambda\mu}(\vartheta,\varphi)] = \sum_{\lambda=2}^{\lambda_{max}} \sum_{\mu=-\lambda}^{\lambda} \alpha^{\star}_{\lambda\mu} Y_{\lambda\mu}(\vartheta,\varphi). \tag{5}$$

[b]Let us remind the reader that the operations called *improper rotations* about a certain axis $\hat{\mathcal{O}}$ through an angle are composed of the usual, i.e., proper rotations followed by a reflection in a plane perpendicular with respect to $\hat{\mathcal{O}}$. According to the terminology used in nuclear physics literature, those improper rotations are sometimes called *simplex* operations.

In what follows we will need a representation of the operators $\hat{g} \in \mathbf{G}$ adapted to the action on the spherical harmonics. Here we consider explicitly the inversion, $\hat{\mathcal{C}}_i$, and spatial rotations denoted $R(\Omega)$. In the latter expression $\Omega$ represents the set of three Euler angles. The plane reflections can be treated explicitly or, alternatively, with the help of the other two operations by employing the group multiplication properties. The simultaneous action of inversion and proper rotations as well as the individual actions of these two operations can be conveniently written down using an auxiliary parameter $\eta$ taking possibly the values 0 (no inversion involved) or 1 (inversion involved):

$$\hat{g} \to \hat{g}(\eta,\Omega) = (\hat{C}_i)^{\eta} R(\Omega). \tag{6}$$

With this notation

$$\text{Eq. (5)}: \ \text{l.h.s.} = \sum_{\lambda=2}^{\lambda_{max}} \sum_{\mu=-\lambda}^{\lambda} \alpha^{\star}_{\lambda\mu} (-1)^{\eta\lambda} \sum_{\mu'} D^{\lambda}_{\mu'\mu}(\Omega) \, Y_{\lambda\mu}(\vartheta,\varphi). \tag{7}$$

Introducing relation (7) into Eq. (5), and re-ordering terms, we find the following equation,

$$\sum_{\mu'=-\lambda}^{\lambda} \sum_{\lambda=2}^{\lambda_{max}} \left[ \sum_{\mu=-\lambda}^{\lambda} \alpha^{\star}_{\lambda\mu} (-1)^{\eta\lambda} D^{\lambda}_{\mu'\mu}(\Omega) - \alpha^{\star}_{\lambda\mu'} \right] Y_{\lambda\mu}(\vartheta,\varphi), \tag{8}$$

that must hold for any pair of angles $\vartheta$ and $\varphi$. Since spherical harmonics are linearly independent, Eq. (8) splits into a system of linear algebraic equations,

$$\sum_{\mu=-\lambda}^{\lambda} \alpha^{\star}_{\lambda\mu} (-1)^{\eta\lambda} D^{\lambda}_{\mu'\mu}(\Omega) - \alpha^{\star}_{\lambda\mu'} = 0. \tag{9}$$

This uniform system of equations can be re-written in a more compact way as

$$\sum_{\mu=-\lambda}^{\lambda} \left[ (-1)^{\eta\lambda} D^{\lambda}_{\mu'\mu}(\Omega) - \delta_{\mu\mu'} \right] \alpha^{\star}_{\lambda\mu} = 0, \tag{10}$$

where $\Omega$ is a fixed set of Euler angles corresponding to a given rotation as defined by $\hat{g}$. For instance in the case of a four-fold $\mathcal{O}_z$-axis rotation this could imply $\Omega = \{\pi/2, 0, 0\}$. The form of the invariance condition in Eq. (10) suggests that solutions can be taken as eigen-vectors of the $(2\lambda+1) \times (2\lambda+1)$ matrix of the form $(-1)^{\eta\lambda} D^{\lambda}(\Omega)$ with the eigen-value equal $+1$, leading to the simplest solution. With the corresponding set of the solutions, say $\bar{\alpha}_{\lambda\mu}$, inserted into (10) we can go back to Eq. (5) to convince ourselves[c] that the condition of invariance under $\hat{g} = \hat{g}(\eta, \Omega)$ is indeed satisfied.

Some remarks may be appropriate at this point. Firstly, solutions corresponding to other eigen-values can be equally acceptable, although the preference can be given to those involving the minimum of non-zero components in terms of $\bar{\alpha}_{\lambda\mu}$. Secondly, because the system of equations in (10) is uniform, multiplying the corresponding vector by a constant corresponds again to a solution. This allows to select, e.g., $\bar{\alpha}_{\lambda\mu=0}$ as an independent parameter, which uniquely fixes all the other non-zero components. By exploring all possible values, say, $\alpha_{\lambda 0}^{min} \le \bar{\alpha}_{\lambda 0} \le \alpha_{\lambda 0}^{max}$ we explore all possible surfaces invariant under the symmetry element $\hat{g}$. Thirdly, all other eigen-solutions correspond to equivalent orientations of the surface under considerations. Fourthly, the number of non-null eigenvectors gives the number of possible orientations.

So far we have presented the solution of a limited problem, i.e., the one of invariance with respect to a single symmetry operation. Formally our problem consists in searching for the *simultaneous* invariance conditions with respect to *all the symmetry elements* $\hat{g} \in \mathbf{G}$. Suppose that there are $f$ elements in the group considered, in which case $\eta_k$ and $\Omega_k$, for $k = 1, 2, \ldots, f$ enumerate the corresponding transformations. In this case, we obtain a system of $f \times \lambda(\lambda + 1)$ equations of the form

$$\left.\begin{array}{rl} \sum_{\mu=-\lambda}^{\lambda} \left[(-1)^{\eta_1 \lambda} D^{\lambda}_{\mu'\mu}(\Omega_1) - \delta_{\mu\mu'}\right] \alpha^{\star}_{\lambda\mu} &= 0 \\ \sum_{\mu=-\lambda}^{\lambda} \left[(-1)^{\eta_2 \lambda} D^{\lambda}_{\mu'\mu}(\Omega_2) - \delta_{\mu\mu'}\right] \alpha^{\star}_{\lambda\mu} &= 0 \\ \ldots &= \ldots \\ \sum_{\mu=-\lambda}^{\lambda} \left[(-1)^{\eta_f \lambda} D^{\lambda}_{\mu'\mu}(\Omega_f) - \delta_{\mu\mu'}\right] \alpha^{\star}_{\lambda\mu} &= 0 \end{array}\right\} . \quad (11)$$

This system of equations is over-defined – it contains more equations than unknowns. The fact that solutions exist (sometimes) is a result of symmetry; this and related problems are a subject of a forthcoming publication. In a practical treatment of the problem we do not need to solve all of these equations. One can use the fact that all the elements of any point-group can be generated out of two, possibly three chosen elements called generators. The original system of $f \times (2\lambda + 1)$ equations reduces to a system of much smaller number of them, yet the property of being over-defined remains for all the groups of interest.

[c] In principle one may expect that under some circumstances (specific choice of the symmetry group $\mathbf{G}$ and of the multipolarity of the class of shapes) there exist degenerate solutions to our problem in (10). In such a case we should be able to construct the same symmetry surfaces by taking arbitrary linear combinations of the degenerate eigen-vectors. This type of situation, however, was not encountered so far in the applications of interest in this article and will not be treated further here.

Finally, let us observe the following mathematical subtlety. Suppose that an operator $\hat{S}$ belongs to the ensemble of the symmetry operations, and thus its action on the surface $\Sigma$ transforms this surface into itself. However, this operation influences also the *rotation-axes* of the rotational-symmetry elements of the group considered – the axes following the operation $\hat{S}$. As a consequence the original group $\mathbf{G}$ ceases being a group of symmetry of the considered surface and it is the new group, $\hat{S}\mathbf{G}\hat{S}^{-1}$, isomorphic with the previous one, that overtakes the invariance rules. This mathematical subtlety has no influence on the physical consequences neither does it influence the interpretation of the discussed symmetry relations, yet it may (and often does) lead to a non-unique description of geometrically/physically equivalent objects, occasionally implying some confusion related to the particular combinations of the three Euler angles and/or signs of some of them.

In the present context we may also arrive at yet another type of technical complications. It is often convenient to limit the parametrization of the surfaces in e.g. deformed Woods-Saxon potential and/or that of the multipole moments in the constrained Hartree-Fock method to *real* $\alpha_{\lambda\mu}$-parameters and/or *real* $Q_{\lambda\mu}$-moments. It may also be convenient to work with a particular spatial representation of the symmetry operators through selection of certain reference frames. Moreover, there exist several ways of selecting the generators of the groups in question, leading to totally equivalent realizations of the group considered, yet possibly differing on the level of constructing the operators within a preselected basis. For instance, consider surfaces generated by combination $(Y_{3,+2} + Y_{3,-2})$ that are invariant under the group realization, say $T_d'$, with the generators selected as $\{C_i \circ C_{4z},\, C_{8z} \circ (C_i \circ C_{4y}) \circ C_{8z}^{-1}\}$. Consider next the surfaces with octahedral symmetry generated by Eqs. (15-17), the corresponding group $O_h'$ generated e.g. by $\{C_i,\, C_{8z}C_{4z}C_{8z}^{-1},\, C_{8z}C_{4y}C_{8z}^{-1}\}$ The latter group contains as its sub-group tetrahedral group $T_d'$ introduced above. The generators of the latter (octahedral) group could have been selected differently, in which case relations (12-14) and (15-18) would have lost consistency.

## 5. Interplay Between Tetrahedral and Octahedral Symmetries

The nuclear mean-field realization, using the parametrized potentials as presented in Sect. 3, is a perfect tool to study the point-group symmetries under the condition of using the *multipole shape parametrization* of Eq. (3). This will become evident from the following considerations.

Our method allows to find the representation of surfaces of a given predefined point-group symmetry through the exact diagonalisation of the Wigner $D$-functions as described in the preceding Section. Such diagonalizations have been performed using *Mathematica* for $\lambda \leq 10$. The first three lowest-order solutions are as follows.

### 5.1. *Tetrahedral Symmetry*

The only solutions that are obtained in terms of the spherical harmonics for $\lambda \leq 10$ are of the third, seventh and ninth order. For each multipolarity we find one inde-pendent deformation parameter that characterizes/defines all the other intervening components. We denote those independent parameters $t_3$, $t_7$ and $t_9$. In the lowest order ($\lambda = 3$) we find only two related spherical harmonics intervening, viz. the ones with $\lambda = 3$ and $\mu = \pm 2$:

$$\alpha_{3,\pm 2} \equiv t_3. \tag{12}$$

We find no solutions of order $\lambda = 5$. In the $\lambda = 7$ order we find four intervening spherical harmonics, i.e., the ones corresponding to $\lambda = 7$, $\mu = \pm 2$ and $\mu = \pm 6$:

$$\alpha_{7,\pm 2} \equiv t_7 \quad \text{and} \quad \alpha_{7,\pm 6} \equiv -\sqrt{\frac{11}{13}} t_7. \tag{13}$$

Finally, for $\lambda = 9$ we obtain

$$\alpha_{9,\pm 2} \equiv t_9 \quad \text{and} \quad \alpha_{9,\pm 6} \equiv +\sqrt{\frac{13}{3}} t_9. \tag{14}$$

We may conclude that there exist only very few spherical harmonics of order $\lambda \leq 10$ that may be used to construct the surfaces of tetrahedral symmetry; but even those that are allowed to intervene are strongly correlated and we have merely 3 independent deformation parameters that characterize the full parametric freedom within tetrahedral symmetry up to the $10^{th}$ order.

### 5.2. *Octahedral Symmetry*

The lowest-order combinations of spherical harmonics that are compatible with the octahedral symmetry of the related surfaces in Eq. (3) correspond to $\lambda = 4, 6, 8$ and 10. Again, within each multipolarity there is only one degree of freedom (one single parameter) that determines the allowed combinations of the allowed spherical harmonics. We denote those parameters, treated as independent, by $o_4$, $o_6$, and $o_8$. The three lowest-order solutions are:

$$\alpha_{4,0} \equiv o_4 \quad \text{and} \quad \alpha_{4,\pm 4} \equiv -\sqrt{\frac{5}{14}}\, o_4 \tag{15}$$

for $\lambda = 4$,

$$\alpha_{6,0} \equiv o_6 \quad \text{and} \quad \alpha_{6,\pm 4} \equiv +\sqrt{\frac{7}{2}}\, o_6 \tag{16}$$

for $\lambda = 6$, and

$$\alpha_{8,0} \equiv o_8, \quad \alpha_{8,\pm 4} \equiv -\sqrt{\frac{28}{198}}\, o_8, \quad \text{and} \quad \alpha_{8,\pm 8} \equiv +\sqrt{\frac{65}{198}}\, o_8 \tag{17}$$

for $\lambda = 8$.

In the following we are going to limit ourselves to the lowest-order degrees of freedom.

### 5.3. *Tetrahedral and Octahedral Degrees of Freedom Combined*

Before proceeding, let us recall in passing an important mathematical relation between the tetrahedral and octahedral symmetries as represented in terms of spherical harmonics. This relation originates from the fact that the tetrahedral symmetry point-group is a sub-group of the octahedral one. Indeed, a surface with the octahedral symmetry is invariant under 48 symmetry elements, among others the inversion. It turns out that the ensemble of the 24 symmetry operations of the octahedral group that do not contain the inversion operation coincides with the 24 symmetry elements of the tetrahedral group. Consequently, all the surfaces invariant under the octahedral symmetry group are at the same time invariant under the tetrahedral symmetry group. One may show that the mathematical expressions given in (12)-(17) are compatible in this sense, and thus an arbitrary combination of nuclear shapes defined by $t_3$, $t_7$, and $t_9$ on the one hand, and $o_4$, $o_6$, and $o_8$ on the other hand, preserves the *tetrahedral symmetry* while setting $t_3 = 0$, $t_7 = 0$ and $t_9 = 0$ we obtain surfaces of pure octahedral symmetry.

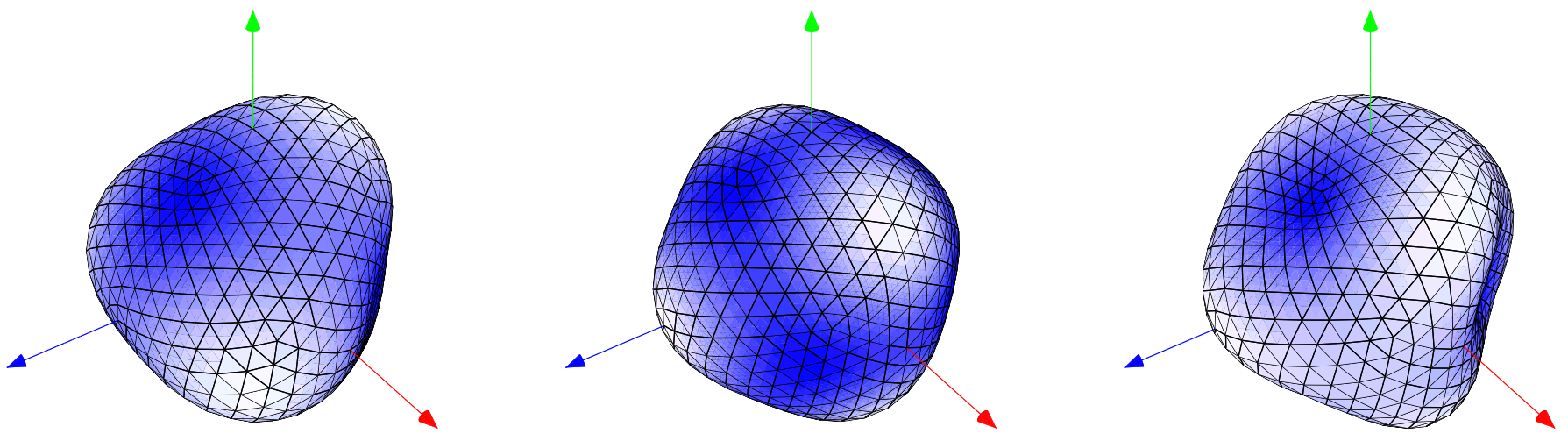

Fig. 1. Illustration of the interplay between the *tetrahedral* and *octahedral* geometrical symmetries. The figure shows a pure tetrahedral symmetry surface corresponding to the deformation $t_3 = 0.15$ (left), compared to the pure octahedral symmetry surface with $o_4 = 0.15$ (middle), compared to the surface obtained by superposition of the two (right). One can demonstrate that the latter surface is still tetrahedrally symmetric. This is also why combining the two symmetries simultaneously may strengthen the final *tetrahedral* symmetry effect.

The above inter-relations are illustrated in Fig. 1 using typical sizes of the tetrahedral and octahedral deformations as predicted by microscopic calculations.

## 6. Tetrahedral Magic Numbers in Rare Earth and Actinide Nuclei

In principle, in the present article we focus our illustrations of the tetrahedral symmetry on the example of the Rare Earth nuclei, but as a 'by-product' the tetrahedral magic numbers in the Actinide nuclei are obtained as well. We begin by the illustration of the realistic single-particle spectra in function of the tetrahedral deformation as calculated for the Rare-Earth nuclei. This choice was dictated by the recent study[9] suggesting that the phenomenon of the tetrahedral symmetry may be present in the whole range of Rare Earth nuclei, in particular in the isotones of $N \sim 90$. Moreover, it seems today[9] that the experimental evidence for the presence of such a symmetry has already been found[d].

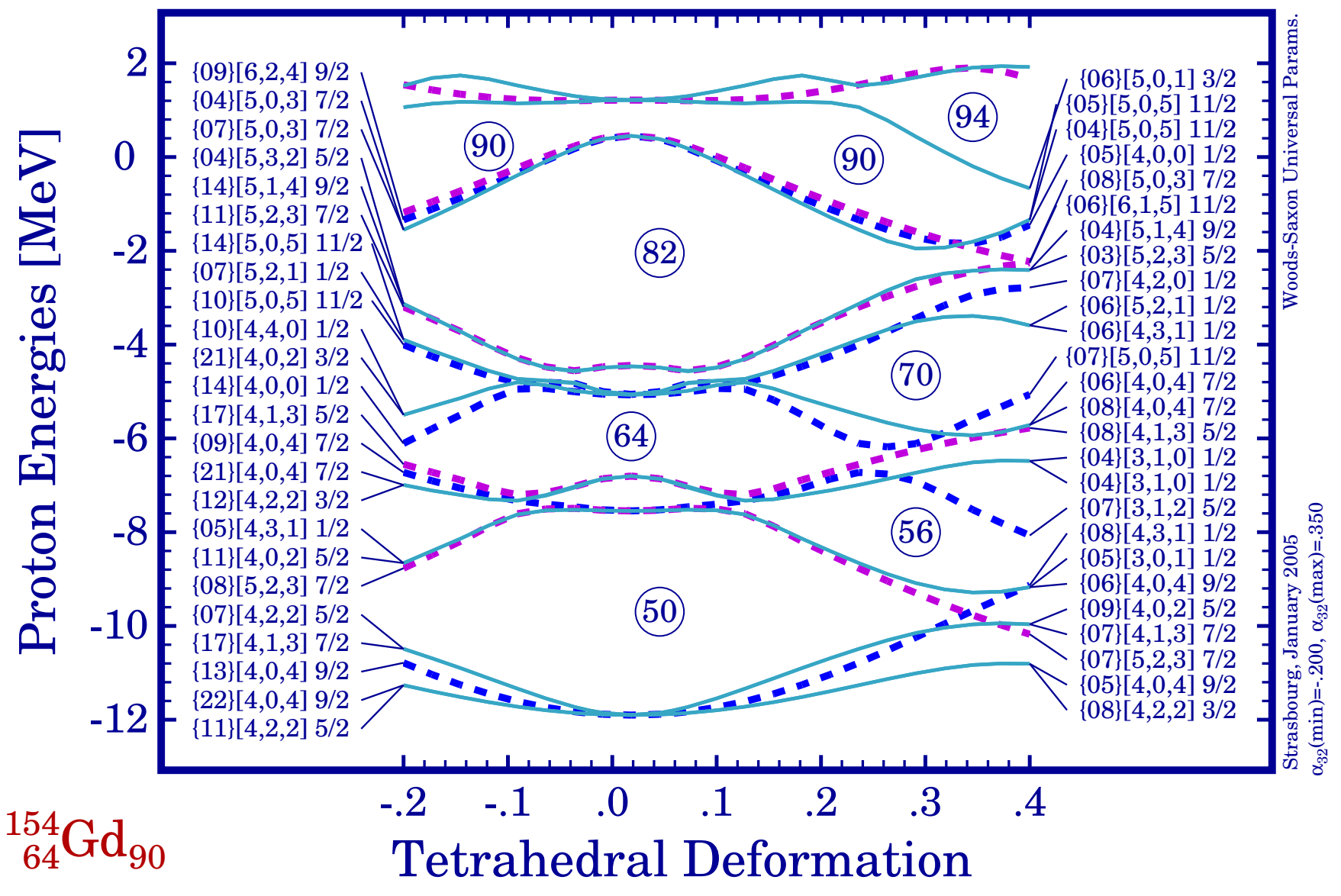

Fig. 2. Single-particle energy levels in function of the tetrahedral deformation for protons. Observe the presence of the four-fold degenerate orbitals marked with full lines and the strong gaps at Z=56, 64, 70, and 90/94 referred to as tetrahedral-magic.

## 6.1. *Single-Nucleon Diagrams*

An illustration of the single-particle proton-spectra for the nucleus $^{154}$Gd is shown in Fig. 2. The presence of the important (~2 MeV) tetrahedral-deformed gaps deserves noticing. These gaps correspond to tetrahedral 'magic' configurations or tetrahedral 'shell closures' at Z=56, 70, and 90/94. In the latter case a huge, about 3 MeV gap is crossed by a single level but the calculated effects of this structure are large; the related discussion will be presented elsewhere. Observe also the four-fold degenerate levels marked as continuous lines. These are the ones corresponding to the four-dimensional irreducible representations of the underlying double tetrahedral point group $T_d^D$. Referring to the same diagram let us emphasize that the spherical $Z = 64$ gap is unstable: the corresponding gap *increases* with the increasing tetrahedral deformation.

[d]The final evidence is expected from comparison of the unique features of the branching ratios of the corresponding electromagnetic transitions. A dedicated study is in progress.

The latter point brings us to a comparison between the spherical and tetrahedral 'magic' numbers. The discussion of *tetrahedral magicity* is a bit more complex as compared to the well known discussions of the spin-orbit splitting and the so-called spherical magic numbers 8, 20, 28, 50, 82 and 126. Indeed, the underlying physics arguments behind the *spherical magicity* are related to the strong main $N$-shell grouping of the single-particle levels together with the intervention of the 'unnatural' parity, highest $[j = (N+1) + \frac{1}{2}]$-orbitals. The presence of these highest-$j$ (intruder) orbitals within the natural parity shells is caused by the strong spin-orbit interaction that is held responsible for such a strong intruder level repulsion and the corresponding spherical gaps.

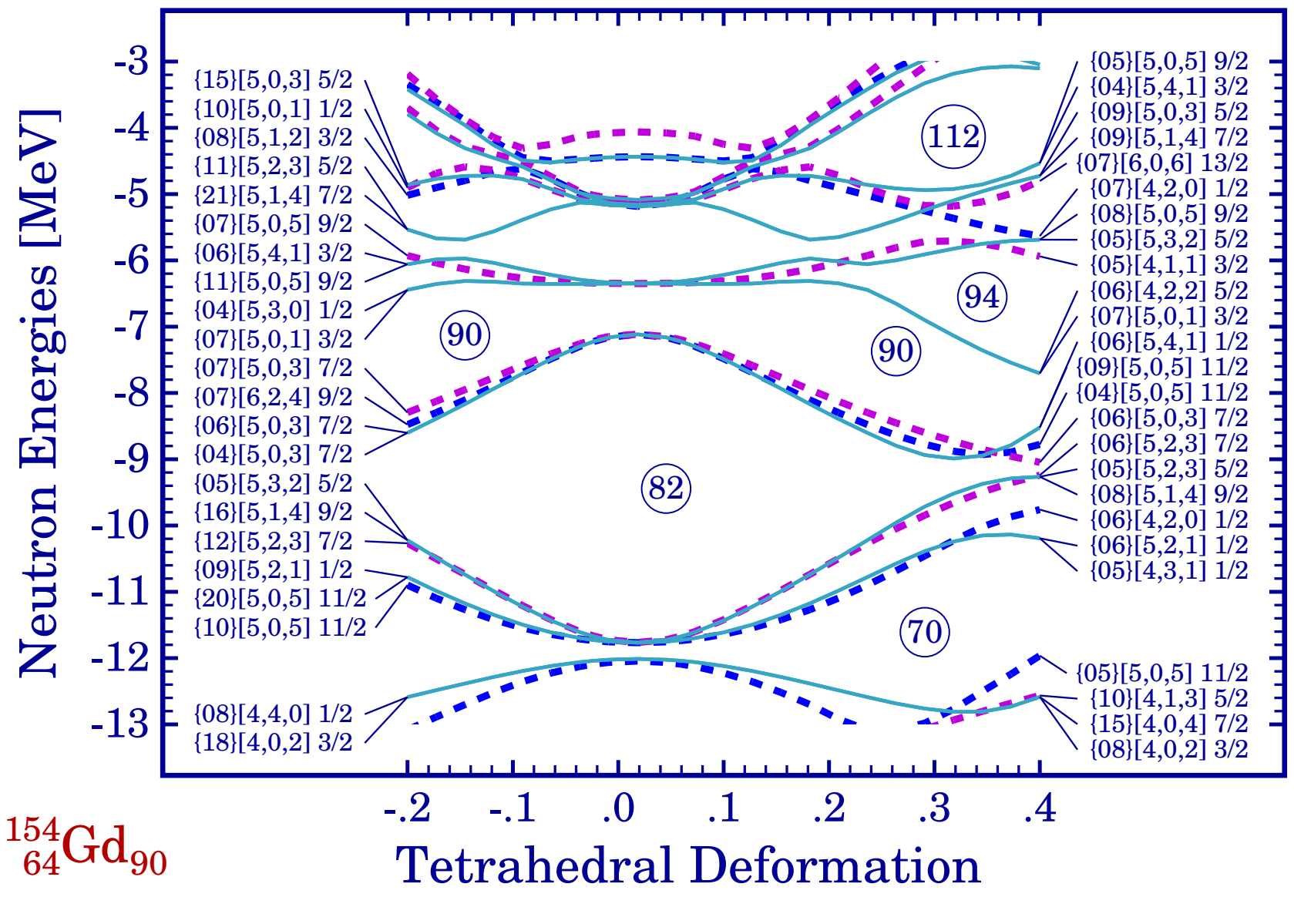

Fig. 3. Single-particle energy levels in function of the tetrahedral deformation for neutrons. Observe strong gaps at N=70, 90/94, and 112.

The calculated tetrahedral symmetry minima on total potential energy surfaces (discussed in the next Section) are a consequence of the spontaneous-symmetry breaking mechanism as the result of which the spherical mean field looses its stability in competition with the intrinsic-parity breaking deformations. The conse-quences thereof for the single-particle spectra is illustrated for the case of neutrons in Fig. 3, analogous to Fig. 2. It is worth emphasizing that magic tetrahedral gaps correspond to the same closures as in the case of protons; even the large gap structures at N=90/94 crossed by a single level look nearly the same as in the case of Z=90/94 configuration. We wish to note the presence of the $N = 112$ gap whose size is comparable to the other tetrahedral gaps in the Figure.

### 6.2. *Tetrahedral Symmetry vs. Hypothetical Neutron-Skin Effect*

The lowest rank tetrahedral deformation corresponds to $t_3 \equiv \alpha_{32}$, an example of octupole deformations. The influence of the octupole degrees of freedom on the single-nucleonic spectra comes mainly through a repulsion between the orbitals that differ in terms of the angular-momentum quantum number $\ell$ by $\Delta\ell = 3$. This repulsion is caused by the presence of the $Y_{\lambda=3}$ components in the mean field and is related to the Clebsch-Gordan coupling in the matrix elements of the type $\langle \ell' | Y_{\lambda=3} | \ell \rangle$. This effect increases with increasing $\ell$ quantum-number of the nucleonic orbitals and thus its influence on the intruder orbitals is the strongest.

Examples are the repulsion effects between $i_{\frac{13}{2}}$-$f_{\frac{7}{2}}$, $h_{\frac{11}{2}}$-$d_{\frac{5}{2}}$ or $g_{\frac{9}{2}}$-$p_{\frac{3}{2}}$, i.e., between the intruder orbitals and their $\Delta\ell = 3$ partners, leading in all cases to an increase of the energy of the *intruder level* and at the same time an increase of the energy spacing between the doublets mentioned when the tetrahedral deformation increases.

Incidentally, there may exist a mechanism in exotic neutron-rich nuclei that influences the behaviour of the intruder orbitals in a very similar way. This mechanism, claimed to possibly occur in very neutron-deficient nuclei, is a hypothetical increasing of the neutron skin[14]. It is equivalent to an effective increase of the diffusivity parameter in the underlying Woods-Saxon potential. Indeed, an elementary estimate for the spherical Woods-Saxon potentials gives

$$V_{WS} \sim \frac{1}{1 + e^{\frac{r-R_0}{a}}} \quad \to \quad V_{so} \sim \left(\frac{\lambda}{a}\right) \cdot \frac{\vec{\ell} \cdot \vec{s}}{\cosh^2(\frac{r-R_0}{2a})}. \tag{18}$$

As can be seen from the above relation an *increase* of the diffusivity parameter $a$ by a certain fraction will immediately lead to a *decrease* of the spin-orbit matrix elements by a similar fraction; this decrease will be furthermore strengthened through the hyperbolic cosine function that is dependent even more strongly on the diffusivity factor $2a$. As a consequence, the skin mechanism, if indeed present in exotic nuclei, will diminish the importance of the spin-orbit potential thus bringing the intruder orbitals higher in energy – in qualitative agreement with the tendency predicted for the tetrahedral deformation effect.

In other words: should the neutron skin effect be discovered in exotic nuclei its presence may induce/strengthen the tetrahedral instability in those nuclei.

## 7. Tetrahedral and Octahedral Symmetry Instabilities

A tendency of generating local minima at non-zero tetrahedral deformation is referred to as the tetrahedral symmetry instability. The corresponding minima are separated with relatively high barriers from the competing, e.g., ground-state minima. It is one of the characteristic features of this mechanism that it *is not* limited to the tetrahedral doubly-magic nuclei. To the contrary, similar effects are predicted for many nuclei in the vicinity of those doubly magic ones.

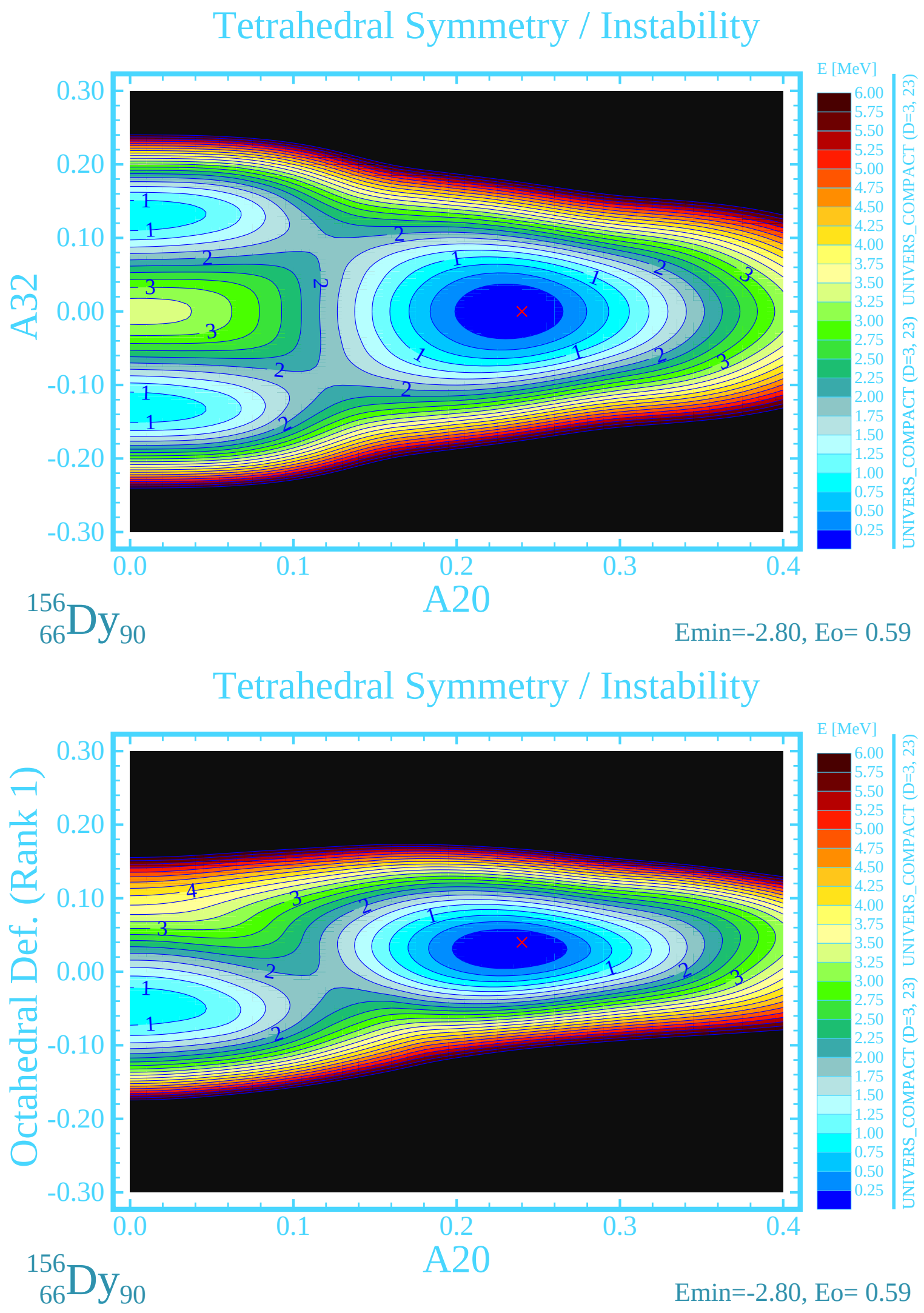

Fig. 4. Total energies for $^{156}$Dy nucleus. Top: in function of the quadrupole ($A20 \leftrightarrow \alpha_{20}$) and tetrahedral ($A32 \leftrightarrow \alpha_{32}$) deformations showing a minimum at about 1 MeV above the prolate deformed ground-state. Bottom: similar in function of the quadrupole deformation and the octahedral deformation of the first rank $o_4$. Observe the $\sim$1 MeV barrier separating the two local minima from the ground-state minimum.

To illustrate this kind of persistent tetrahedral effects, we have chosen the $^{156}$Dy nucleus that has 2 protons in excess of the tetrahedral magic $Z = 64$ gap. The corresponding illustration in Fig. 4 was obtained by minimizing the total energy in the 3-dimensional deformation space $(\alpha_{20}, t_3 \equiv \alpha_{32}, o_4)$. Each map was constructed by projecting the total energy onto the variable that is not marked on the x- and y-axis ($o_4$ for the top frame, $t_3$ for the bottome frame). The double minimum structure in the deformation plane $\alpha_{20}$-$\alpha_{32}$ (quadrupole-tetrahedral) is clearly visible. Comparisons show that the tetrahedral deformation brings over 3 MeV of energy gain in this nucleus (as compared to the original energy at the spherical shape). Similarly, the octahedral deformation brings an additional energy gain of about 0.5 MeV. Thus both types of symmetries combine to creating a final minimum with *tetrahedral symmetry only*, made of the superposition of *pure tetrahedral* and *pure octahedral* symmetry components of the nuclear surfaces.

So far we have presented the results based on the calculations employing the non-selfconsistent deformed Woods-Saxon potential; it will be instructive to verify the predictions employing the self-consistent Hartree-Fock method as presented in the next Section.

## 8. Tetrahedral and Octahedral Instabilities within Self-Consistent HFB Method

It is worth emphasizing that self-consistent solutions also clearly manifest the presence of the tetrahedral symmetry. Table 1 shows the results of the Hartree-Fock-Bogolyubov (HFB) calculations performed for several Rare-Earth nuclei by using the SIII[15] parametrization of the Skyrme force and the zero-range density-dependent mixed pairing force[16,17]. We used the code HFODD (v2.20m)[18,19,20] to solve the self-consistent equations on the basis of $N_0 = 16$ spherical harmonic-oscillator shells.

Contrary to non-selfconsistent methods, in the self-consistent calculations nuclear deformations are not parameters of the potential but result from dynamical effects related to the self-optimization of the nuclear shape. Therefore, the values of the multipole moments of the mass distribution, shown in Table 1, are the results of the calculation and not subject to any pre-assumption. The results shown in column no. 7 of the Table represent $Q_{40} \times \sqrt{\frac{5}{14}}$. The latter expression, according to Eq. (15), should be equal to $-Q_{44}$ if the solution possesses the octahedral symmetry. This relation is verified with a very high precision, as the comparison of columns 6 and 7 shows.

The solutions presented point to the presence of the tetrahedral instability around $Z = 64$ and $N = 90$ nuclei that may amount to about $-4\,$MeV as in the case of $^{156}$Gd. This latter case deserves particular attention since the corresponding solution is characteristic for its vanishing $Q_{32}$ moment and can be seen as an example of *pure octahedral symmetry.* However, in most cases we have at the same time the tetrahedral moment $Q_{32} \neq 0$ and the hexadecapole moments $Q_{40} \neq 0$ and $Q_{44} \neq 0$, the latter appearing in the exact proportions characteristic for the octahedral symmetry.

Table 1. The HFB results for tetrahedral solutions in light Rare-Earth nuclei. The third column gives the energy difference between the spherical and the tetrahedral minima; for instance the energy difference $\Delta E = -1.387$ implies that the tetrahedral minimum lies 1.387 MeV below the corresponding energy of the spherical configuration. Columns 4, 5, and 6 give the multipole moments indicated. For comments concerning column 7 – see text.

| Z | N | $\Delta E$ (MeV) | $Q_{32}$ ($b^{3/2}$) | $Q_{40}$ ($b^2$) | $Q_{44}$ ($b^2$) | $Q_{40} \times \sqrt{\frac{5}{14}}$ ($b^2$) |
|---|---|---|---|---|---|---|
| 64 | 86 | −1.387 | 0.941817 | −0.227371 | +0.135878 | −0.135880 |
| 64 | 90 | −3.413 | 1.394656 | −0.428250 | +0.255929 | −0.255928 |
| 64 | 92 | −3.972 | 0.000000 | −0.447215 | +0.267263 | −0.267262 |
| 62 | 86 | −0.125 | 0.487392 | −0.086941 | +0.051954 | −0.051957 |
| 62 | 88 | −0.524 | 0.812103 | −0.218809 | +0.130760 | −0.130763 |
| 62 | 90 | −1.168 | 1.206017 | −0.380334 | +0.227293 | −0.227293 |

Although the detailed properties of these solutions depend quite strongly on the parametrization of the Skyrme- and pairing interaction, the results presented here show all the characteristic features of the exotic symmetries as discussed earlier in the framework of the non self-consistent approaches. In particular, it illustrates very clearly that the combination of the tetrahedral and octahedral symmetries can lower the total energy of the system leading to the final *tetrahedral symmetry* as discussed earlier on the basis of the group-theoretical considerations in Sect. 5.

## 9. Summary and Conclusions

We have suggested that the new approach to the problem of the nuclear stability, in particular for the exotic nuclei projects, should take advantage the point group symmetries: especially those groups that possess large number of irreducible representations seem most promising. We have discussed the mechanism of high-rank geometrical symmetries, tetrahedral and octahedral ones, within the nuclear mean-field approach. We have presented in some detail how the conditions of invariance of the nuclear surfaces against the symmetry operations of a given symmetry group **G** can be implemented when using the spherical harmonics representation of the nuclear surfaces.

It turns out that the superposition of tetrahedral and octahedral nuclear surfaces still lead to a tetrahedrally-invariant deformed Woods-Saxon mean-field Hamiltonian. The same argument applies of course also for the self-consistent Hartree-Fock calculations as explicitly demonstrated through calculations for several nuclei in the considered range of the light Rare-Earth nuclei. Calculations with the Woods-Saxon Hamiltonian show that by combining the two symmetries an extra gain of up to about 1.5 MeV can be achieved. Illustrations related to the tetrahedral magic numbers and of the total potential energy surfaces have also been presented, fully confirming the general group-theoretical considerations as well as qualitative considerations related to the density of the nucleonic levels.

## ACKNOWLEDGMENT

This work was partially supported through the collaboration program between the $IN_2P_3$, France, and the Polish partner Institutions; by the Polish Committee for Scientific Research (KBN) under Contract No. 1 P03B 059 27, and by the Foundation for Polish Science (FNP).

## References

1. I. Hamamoto, B. Mottelson, H. Xie, and X. Z. Zhang, Z. Phys. **D21**, 163 (1991).
2. X. Li and J. Dudek, Phys. Rev. **C49**, R1250 (1994).
3. S. Takami, K. Yabana, and M. Matsuo, Phys. Lett. **B431**, 242 (1998);
   M. Yamagami and K. Matsuyanagi, Nucl. Phys. **A672**, 123 (2000);
   M. Yamagami and K. Matsuyanagi, and M. Matsuo, ibid. **A693**, 579 (2001)
4. J. Dudek, A. Góźdź, N. Schunck, and M. Miśkiewicz,
   *Phys. Rev. Lett.*, **88**, 252502 (2002).
5. J. Dudek, A. Góźdź, and D. Rosły, Acta Phys. Polonica, **B32**, 2625 (2001)
6. A. Góźdź, J. Dudek, and M. Miśkiewicz, Acta Phys. Polon. **B34**, (2003) 2123
7. J. Dudek, A. Góźdź, and N. Schunck, Acta Phys.Polon. **B34**, 2491 (2003);
8. N. Schunck, J. Dudek, A. Góźdź, and P.H. Regan, Phys. Rev. **C69**, R061305 (2004).
9. J. Dudek, D. Curien, N. Dubray, J. Dobaczewski, V. Pangon, P. Olbratowski, and N. Schunck; *Phys. Rev. Lett.* **97**, 072501 (2006)
10. P. Olbratowski, J. Dobaczewski, P. Powałowski, M. Sadziak, and K. Zberecki,
    *Int. J. Mod. Phys.* **E13**, 333 (2006).
11. K. Zberecki, P. Magierski, P.-H. Heenen, and N. Schunck,
    Phys. Rev. C (2006) *in press*, nucl-th/0604047.
12. T. R. Werner and J. Dudek, Atomic Data Nucl. Data Tables **50**, 179 (1995);
    T. R. Werner and J. Dudek, Atom. Data Nucl. Data. Tables **59**, 1 (1995)
13. J. Dudek, Proc. Int. Winter Meeting on Nuclear Physics, Bormio, Italy 1987;
    *(Ricerca Scientifica ed Educazione Permanente, Supplemento N.56*, Edited by I. Iori)
14. J. Dobaczewski, I. Hamamoto, W. Nazarewicz and J.A. Sheikh,
    Phys. Rev. Lett. **72**, 981 (1994)
15. M. Beiner, H. Flocard, N. Van Giai and P. Quentin, Nucl. Phys. **A238**, 29 (1975)
16. J. Dobaczewski, W. Nazarewicz and M.V. Stoitsov,
    Proceedings of the NATO Advanced Research Workshop *The Nuclear Many-Body Problem 2001*, Brijuni, Croatia, June 2-5, 2001,
    eds. W. Nazarewicz and D. Vretenar (Kluwer, Dordrecht, 2002), p. 181
17. J. Dobaczewski, W. Nazarewicz, and M. V. Stoitsov, Eur. Phys. J. A **15**, 21 (2002)

18. J. Dobaczewski and P. Olbratowski, Comput. Phys. Commun. **158**, 158 (2004); **167**, 214 (2005)
19. J. Dobaczewski *et al.*, HFODD home page: `http://www.fuw.edu.pl/~dobaczew/hfodd/hfodd.html`
20. J. Dobaczewski *et al.*, Comput. Phys. Commun., to be published